\documentclass{article}

\usepackage{epsfig}

\begin{document}

\begin{center}

{\Large\bf Lattice QCD as a video game}

\vspace{1cm}
Gy\H oz\H o I. Egri$^{\,a}$,\, Zolt\'an Fodor$^{\,abc}$,\, Christian Hoelbling$^{\,b}$, \\
S\'andor D. Katz$^{\,ab}$,\, D\'aniel N\'ogr\'adi$^{\,b}$\, and\, K\'alm\'an K. Szab\'o$^{\,b}$
\vspace{0.5cm}

\end{center}

\hspace{2cm}$^a${\it Institute for Theoretical Physics, E\"otv\"os University, Budapest, Hungary}

\hspace{2cm}$^b${\it Department of Physics, University of Wuppertal, Germany}

\hspace{2cm}$^c${\it Department of Physics, University of California, San Diego, USA}

\vspace{0.5cm}

\abstract{The speed, bandwidth and cost characteristics of today's PC graphics cards 
make them an attractive target as general purpose computational
platforms. High performance can be achieved also for lattice simulations but the actual
implementation can be cumbersome. This paper outlines the architecture and
programming model of modern graphics cards for the lattice practitioner with the goal of
exploiting these chips for Monte Carlo simulations. Sample code is also given.}

\section{Introduction}
\label{introduction}

The goal of every lattice field theorist is to use a calculational platform that maximizes the performance/price ratio. In this
paper a competitive but so far unused and unappreciated (at least in the lattice community) architecture will be introduced.

So far the only available option was the usage of CPU-based platforms may it be individual PCs, PC clusters, dedicated
supercomputers such as QCDOC or APE or commercial supercomputers such as BlueGene/L. The actual calculational
task in all of these solutions is done by CPU's which
significantly vary in terms of features but are similar in the sense that they
all share a very general purpose architecture. In recent years a rapidly developing
specialized architecture emerged from the graphics industry, Graphical Processing Units or GPU's which took over some of the
calculational tasks of the CPU.  These chips are designed to fulfill the needs of a graphics oriented audience (gamers, designers,
etc.) and hence were specialized to the kind of task this set of users most frequently need i.e.  graphical processing. However
the complexity of this task grew to a level that general programmability of the chips was also required. The end product of this
evolution is a high performance chip optimized for SIMD floating point operations on large vectors that can be utilized for
general purpose calculations such as lattice field theory.

\begin{figure}
\label{perf}
\centering
\includegraphics[width=11cm]{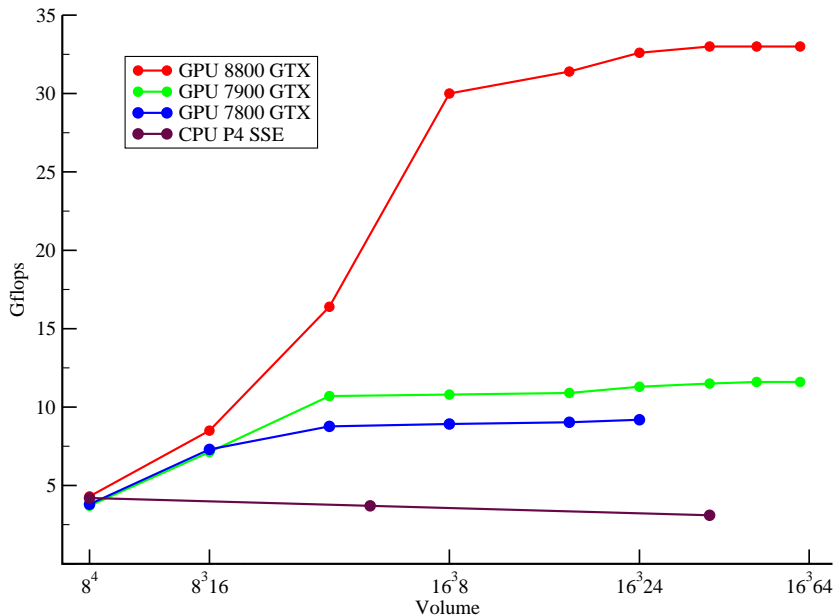}
\caption{Wilson matrix multiplication performance for various lattice volumes on 3 different NVIDIA cards and a 3 GHz Pentium 4 using SSE optimization.}
\end{figure}

Figure 1 and 2 show sustained performances for both Wilson and staggered matrix multiplication on various lattice sizes and a
comparison is given with SSE optimized CPU codes on an Intel P4. Considering the fact that the price of the current top GPU models
are around \$500 it becomes clear that they are very cost effective. For reference we give some numbers from
figure \ref{perf} for the NVIDIA 8800 GTX card: 33 Gflops sustained performance on a $16^3\times60$ lattice using the
Wilson kernel. Another good reason for investigating graphics hardware is the fact that the performance growth rate is still a steep
exponential for GPU's \cite{survey}.

The relatively low price tag of GPU's is of course the result of the large market value of their target audience (gamers,
designers, etc.) which was also the reason why ordinary PCs proved to be very cost effective in the past
\cite{Csikor:1999vz, Fodor:2002zi}.

The question of scalability is of course an important one for any high performance calculational platform and for GPU's this
aspect has not yet been explored in detail for lattice applications. The various possibilities for using multiple GPU's in a parallel computation are
mentioned briefly in section \ref{present} as well as some notable difficulties related to GPU programming. A scalable GPU cluster
for flow simulations has been reported in \cite{gpucluster}.

Much more details on general purpose programming of graphics hardware than what is covered here can be found in \cite{survey}, see
also \cite{gems} as well as the tutorials \cite{tutorial}.

\section{Architecture}
\label{gpuarchitecture}

In order to understand the basics of GPU programming and the techniques of writing efficient code the architecture of modern GPU's
will be described briefly below; for more details see \cite{survey}.

\begin{figure}
\label{xx}
\centering
\includegraphics[width=11cm]{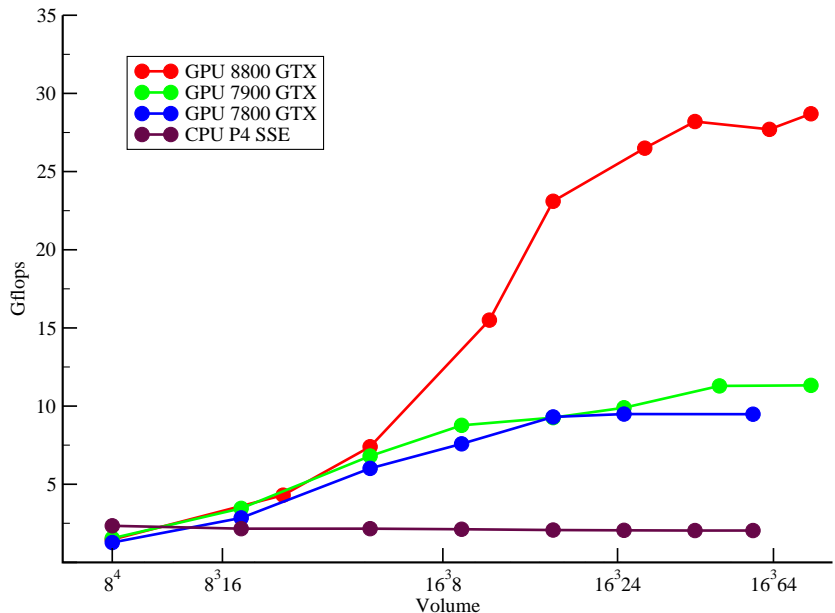}
\caption{Staggered matrix multiplication performance for various lattice volumes on 3 different NVIDIA cards and a 3 GHz Pentium 4 using SSE optimization.}
\end{figure}

The native data type is a 2-dimensional array called a texture. In graphics applications such as games the content of a texture is
typically displayed on the screen after applying a set of transformations on it. On modern cards these transformations can be (almost)
freely programmed which allows us coding for instance lattice applications. Each texture has a width and height and consist of
width $\times$ height number of pixels.  There are various types of textures but in our lattice application we use one which
contains 4 floating point numbers per pixel which are the RGBA (red, green, blue, alpha) color channels\footnote{The alpha color
channel represents transparency in graphics applications}. In some special cases we
also use textures which store only one floating point number per pixel; see section \ref{latticelayout}.

In any calculation a set of textures is used as input and another set of textures are calculated as output. The incoming textures
can be of different sizes whereas the outgoing textures must be of the same size. The incoming textures
are read only while the outgoing textures are write only, thus the calculation must be
such that the outgoing pixels can be calculated independently and can not refer to each other. This constraint allows for
massively parallel execution of the code, i.e. a large number of outgoing pixels are
computed at the same time. This feature together with the two facts that (1) 24 or more
processors are included in a single modern GPU and (2) the availability of a very high memory bandwidth, is the source
of the high performance of current chips. Since the chip acts as a streaming processor by executing instructions in parallel the
architecture is not of the classic von Neumann type.

The question might still arise how it is possible that the growth rate is much higher for GPU's than CPU's since both chips are
the product of the same semiconductor industry. Part of the answer is that in a CPU many transistors are busy with
non-computational tasks such as branch prediction and more importantly caching, etc,
while in a GPU almost all transistors are calculating \cite{survey}.

The task of calculating outgoing textures from incoming textures as outlined above is done by fragment processors.
There are other components in the graphics cards such as vertex processors but we did not find
these useful for our lattice code. Some of the modern chips have unified the tasks of the vertex and fragment
processors thus achieving even more general purpose programmability.

% As reference we list below the specification of the NVIDIA 7900 GTX card: 24 fragment processors, 800MHz clock speed, 8 floating
% point operations per cycle per processor, 512 MB DDR3 memory at 1600 MHz, 256 bit wide bus, 51.2 GB/s peak memory bandwidth. This
% results in 154 Gflops theoretical peak performance (for sustained performance see figures \ref{perf} and \ref{xx}).

As reference we list below the specification of the NVIDIA 8800 GTX card: 128 fragment processors, 1350MHz clock speed,
768 MB DDR3 memory at 1800 MHz, 384 bit wide bus, 86.4 GB/s peak memory bandwidth. 

\section{Programming}
\label{programming}

In order to exploit the unusual but powerful architecture of GPU's a somewhat unusual programming model must be used. In this
section the basics of GPU programming will be introduced with emphasis on the parts that are needed for a lattice application
\footnote{The latest NVIDIA 8800 GTX card supports the so-called Compute Unified Device Architecture (CUDA) which allows for a simpler
programming model but a discussion of this topic is beyond the scope of the present paper \cite{cuda}.}.

The basic software framework for interacting with the GPU from the CPU is OpenGL \cite{opengl}. It is a graphics library
containing functions and macros that can be used to manage textures on the GPU. For instance OpenGL function calls create, delete,
bind, select textures as well as set their properties. The core OpenGL library has many extensions typically invented by the chip
vendors that are thus tailored to a specific hardware. Once such an extension becomes wide spread it may become part of the OpenGL
core specification. The list of general OpenGL extensions can be found in \cite{openglext} while the NVIDIA specific extensions are
available from \cite{NVIDIAopengl}.

Another important segment of GPU programming is the graphics driver that comes with the actual cards and should be provided by the
chip vendors in order to exploit all hardware features. Unfortunately most high quality drivers are closed source. Needless to say
that developers like lattice field theorists using GPU's for general purpose calculations would benefit immensely from some degree
of openness regarding both the hardware and the driver.

The drivers are important because every OpenGL call or any instruction from the CPU is given to the driver which then decides what
exactly is passed to the GPU. For instance it may optimize a set of calls into simpler calls if possible. Usually such an
optimization can make a drastic effect on performance thus the usage of the latest drivers (specific to the given chip) is
recommended.  Since from the two major vendors, ATI and NVIDIA, only NVIDIA offers high quality drivers for Linux we are using
exclusively NVIDIA hardware.

The third and perhaps the most important segment of GPU programming is the actual computation on the textures, i.e. the code that
specifies how a set of outgoing textures are computed from a set of incoming textures. This kind of code is typically called a
pixel shader and can be written in a number of ways. The crudest option is the usage of a GPU-specific assembly-like language. A
higher-level alternative is a C-like language called Cg (C for graphics) \cite{cg} while there are some even higher-level tools
like Brook/BrookGPU \cite{brook} or Sh \cite{sh}. In our lattice code we used Cg.

The available instruction set is a powerful one, for instance there is a single multiply-add instruction on the assembler level
and the shuffling (also called swizzling) of the RGBA components is trivial i.e. an operation like $A.rgba = B.braa * C.rrgg +
D.rgba$
is translated into a single assembler instruction. Here $A$, $B$, $C$ and $D$ represent pixels and for instance $A.rgba$
and $B.braa$ refers to the red, green, blue and alpha values of $A$ and the blue, red, alpha and alpha values of $B$
and similarly for $C.rrgg$ and $D.rgba$. This representation is the actual syntax of Cg.

\section{The lattice on a GPU}
\label{latticelayout}

We currently have production code for both dynamical overlap (Wilson) and staggered fermion formulations. In this section the
basic layout of both formulations will be outlined as well as the details necessary for conjugate gradient algorithms.

\subsection{Wilson layout}
\label{wilsoncode}

If the lattice dimensions are $N_x$, $N_y$, $N_z$ and $N_t$ then each texture will have width $N_x N_y$ and height $N_z N_t$ thus
a lattice site will correspond to a pixel.

The gauge links in a given direction are given by 9 complex or 18 real numbers per site. These 18 numbers fit into 5 textures
since each pixel can store 4 floating point numbers (the RGBA color channels) in the type of textures we use. In fact the 5
textures can store $5\cdot 4 = 20$ numbers so along with the 18 links components there is space for 2 more numbers. These can be used to
store the 2-dimensional texture location of its nearest neighbor in the given direction. Thus over all the gauge links are stored
in 20 textures, 5 for each direction.

A Wilson vector has 4 Dirac and 3 color components per site making up 24 real components. These can be stored in 6 textures.

For sustained performance of our Wilson code see figure \ref{perf}.

\subsection{Staggered layout}
\label{staggeredcode}

In order to exploit the parallelism associated with the 4 RGBA channels we have implemented the staggered code also using this
texture type although it was slightly more complicated than in the Wilson case because staggered fermions only have one complex
component per site per color. Thus two sites (two complex numbers) were stored in an RGBA type pixel and each texture was $N_x N_y / 2
\times N_z N_t$. All further textures used in the staggered code were also of this size.

Each color component of a staggered vector was stored in different textures resulting in 3 textures per vector.

In order to save space and bandwidth only the first two rows of the links are stored in the staggered implementation, the last row is
reconstructed as needed (see below for details). The associated calculational overhead was not included in figure \ref{xx}.
The first two rows give 6 complex numbers per direction per site and are then stored in 6 textures per direction, altogether
giving 24 textures for the links.

In addition an RGBA type index texture is employed per direction to store the needed information for reconstructing the third row of
the links as well as information about neighbors. Out of the 4 RGBA values
red and green store the 2-dimensional texture location of the neighbor and the blue and alpha values store the staggered phase of the
corresponding 2 sites needed for the calculation of the third row of the link matrices.

For sustained performance of our staggered code see figure \ref{xx}.

\subsection{Linear algebra and conjugate gradient}

Once the storage layout is determined as above one only needs to write pixel shaders in Cg to do the actual computations such as
multiplying by the fermion matrix, linear algebra, etc. Since typically the majority of the execution time is spent on inversions
we have implemented both conjugate gradient and multi-shift conjugate gradient algorithms on the GPU.

It must be noted that currently GPU's are only able to handle single precision floating point numbers (unfortunately with non-IEEE
compliant rounding) and the resulting precision might not be
sufficient for a typical conjugate gradient algorithm in QCD. However since the CPU is always available with its double precision
arithmetic units it is possible to use so-called mixed precision inverters. This method consists of inverting in single precision,
calculating the residue in double precision and restarting the single precision inverter from the residue. This iterative
procedure can be repeated any number of times always accumulating the single precision result vectors in double precision until
the final result vector will be correct to double precision. Obviously, the overwhelming majority of matrix multiplications will
be done in single precision on the GPU and only a small number of multiplications will be performed in double precision on the
CPU. For more details of mixed precision interters as well as their implementations on GPU's see \cite{dom}.

As an example we
give concrete numbers for a typical $16^4$ Wilson configuration on the NVIDIA 7900 GTX model: the full double precision inverter on the CPU took 730
iterations to reach a relative $10^{-10}$ precision for the residue while the mixed precision method took 840 single precision
iterations on the GPU and was restarted 4 times, thus there were 4 double precision multiplications. The speed up of the method
was around a factor of 10 which roughly consists of $5 \times 2$ where 5 is coming from the fact that the 7900 GTX model is around
5 times faster than the CPU when both are using single precision arithmetic and the factor of 2 comes from the memory bandwidth
gain of using single and not double precision floating point numbers 99\% of the time. Naturally, a mixed precision technique can
also be implemented fully on the CPU which would speed up the CPU calculation by roughly a factor of 2 leading to an over all
speed up of a factor of 5 for the GPU relative to the CPU.

We have written separate pixel shaders for multiplication by the fermion matrix, linear algebra operations and scalar products.
Except for the latter the graphics architecture can be exploited easily since both matrix multiplication and linear algebra
operations are efficiently parallelizable. However it is less efficient for reducing a vector (or vectors) to a single number and
thus the calculation of scalar products requires some care. 

The most straightforward way is performing all operations at every site such that only a single number remains per site. This
single number per site is stored in a texture that only contains one number per pixel. This texture is then transferred to CPU
memory and is summed there to produce the actual scalar product we were looking for. A more advanced technique we implemented is
using the same method to reduce the scalar product to a single number per site but doing some of the summing on the GPU. For
instance one can divide the texture into 2, 4, 8 or 16 equal size subtextures and sum over these subtextures producing a texture
with the size 1/2, 1/4, 1/8 or 1/16 of the original texture.
This smaller texture can be transferred faster from GPU memory to CPU memory simply
because of the smaller size and the remaining sum can be performed on the CPU. The number of subtextures is a parameter that can
be optimized for a given lattice size and bus speed.

In order to speed up the GPU - CPU transfer rates we have employed a new NVIDIA OpenGL extension called Pixel Buffer Objects or
PBO's \cite{pbo}. This allows for asynchronous readbacks from the GPU and saves one data copy from the driver controlled memory to
the programmer controlled memory. In our lattice code we found it very advantageous to use this new extension because it
considerably increases the GPU - CPU transfer rate.

\section{Present difficulties and future prospects}
\label{present}

There are a number of unpleasant features and limitations at present that unavoidably effect anyone attempting to do lattice
simulations on a GPU.

At present each of our cards is running separately and there is no communication between the nodes. In principle there are two
distinct approaches for introducing communication between cards. First, if the motherboard has two available PCI express slots two
cards might be used in one node. Successfully implementing such a solution would substantially reduce costs of course. A second
and probably simpler option would be installing only one card in one node and have usual communication between nodes (Ethernet,
InfiniBand, etc). We hope to study both of these possibilities in the future with the goal of building a scalable GPU cluster.

A second issue is that the current top NVIDIA model is the 8800 GTX which has only 768 MB memory. This will certainly change in the future.

Regarding precision, most GPU's at present are only capable of handling single precision numbers as was mentioned in the previous
section. It was also detailed how to overcome this limitation in a conjugate gradient algorithm in such a way that 99\% of the
calculation is done in single precision while the end result is correct to double precision.

Since the memory modules of GPU's lack error correction manual checking of the results is recommended as much as this is feasible
similarly to PC's with non-ECC memories. For instance the result vector from a conjugate gradient algorithm should be checked on
the CPU in order to eliminate possible GPU memory failures.

Installing a modern GPU in a PC approximately doubles the power consumption. This fact should not be overlooked when
choosing the appropriate PC hardware hosting the GPU.

It is somewhat of an inconvenience that in most current cards there is very limited support for conditional statements and loops and they drastically
lower the performance. However we did not find this to be a serious limitation as we completely eliminated every
conditional statement and loop from the pixel shaders (Cg code) i.e. from the part that is executed on the GPU.

In practice a further slight annoyance is the lack of simple debugging mechanisms as the programmer does not have direct access to the GPU
memory, only through the driver. It is recommended to use the built-in debugging features of OpenGL and Cg.

Although the learning curve for GPU programming is initially rather steep once the needed basics of OpenGL are absorbed the actual
coding in Cg is not that much different from ordinary C and definitely simpler than hand written (SSE optimized) assembly. The
actual man-time needed to develop GPU lattice gauge theory code of course very much depends on personal skills and
programming background. In general if one starts from scratch it is less effort than a high quality mix of C and optimized assembly but
certainly more effort than using pure C or readily available high-level libraries such as Chroma \cite{chroma}. It is worth noting
that the recently introduced Compute Unified Device Architecture (CUDA) of NVIDIA will clearly reduce the man-time needed for implementation
\cite{cuda}.

As to the future, the hardware, corresponding driver and OpenGL extension improvements on the side of the vendors is rather
unpredictable except for the obvious trend to improve the existing features (more memory, more bandwidth, faster clock speed,
etc.). An example of a useful recently added OpenGL extension is the PBO extension mentioned in the previous chapter.

\section{Sample code}
\label{samplecode}

A full working code that performs the operation $z_i = x_i + y_i$ on the GPU for
three $4NM$ sized single precision arrays is described below. In order to use the code the following software must be installed on a Linux
system: glew, freeglut, Cg and an appropriate driver.

As mentioned in section \ref{programming} each data array $x, y$ and $z$ will be stored in textures. Since their size was chosen
to be $4NM$ the size of the textures will be $N\times M$ and each pixel will store 4 floating point numbers.

Below is a Cg program that adds two textures and puts the result into a third texture. Suppose the following code is saved in a
file called {\tt add.cg}.
\vspace{0.5cm}

{\footnotesize
\begin{verbatim}
/* define a structure corresponding to a single output texture */
struct FragmentOut { float4 color0:COLOR0; };

/* this is the pixel shader doing the addition */
FragmentOut add( in float2 myTexCoord:WPOS, uniform samplerRECT x, uniform samplerRECT y )
{
    FragmentOut c;
    c.color0 = texRECT( x, myTexCoord ) + texRECT( y, myTexCoord );
    return c;
}
\end{verbatim}
}
\vspace{0.5cm}

Notice that the line containing the addition adds 4 floating point numbers in one go since each texture lookup, texRECT( \ldots ),
results in a structure containing all the 4 numbers stored in a pixel.

Once the pixel shader is ready the actual C program containing all OpenGL calls necessary to
create the textures, load the initial data $x$ and $y$ from the CPU memory to the GPU memory, run the above shader and retrieve the result from
the GPU memory back into CPU memory, finally comparing the result of the addition with the same result obtained on the CPU should be
written. This C code is given below (without using PBO's) and it is supposed that it is saved as {\tt add.c}. The comments should
make it clear what the purpose of most OpenGL and Cg calls are.
\vspace{0.5cm}

{\footnotesize
\begin{verbatim}

#include <stdio.h>
#include <stdlib.h>
#include <string.h>
#include <GL/glew.h>
#include <GL/glut.h>
#include <Cg/cg.h>
#include <Cg/cgGL.h>

/* define texture width = N and height = M and vector size = 4*N*M */
#define N 640
#define M 480
#define v 4*N*M

/* we will have only this single function */
int main( int argc, char **argv )
{
    int i;

    /* create three pointers to store our vectors x, y, z */
    float * x = ( float * ) malloc( v * sizeof( float ) );
    float * y = ( float * ) malloc( v * sizeof( float ) );
    float * z = ( float * ) malloc( v * sizeof( float ) );

    /* fill up y and z with some data */
    for ( i = 0; i < v; i++ )
    {
        x[i] = (float)( i / 10.0 );
        y[i] = (float)( i / 20.0 );
    }
    
    /* set up glut and glew */
    glutInit( &argc, argv );
    glutCreateWindow( "add" );
    glewInit(  );
    
    /* some obligatory OpenGL settings */
    glMatrixMode( GL_PROJECTION );
    glLoadIdentity(  );
    gluOrtho2D( 0.0, N, 0.0, M );
    glMatrixMode( GL_MODELVIEW );
    glLoadIdentity(  );
    glViewport( 0, 0, N, M );

    /* set up Cg context and profile */
    CGcontext context = cgCreateContext(  );
    CGprofile profile = cgGLGetLatestProfile( CG_GL_FRAGMENT );
    cgGLSetOptimalOptions( profile );
   
    /* create fragment program 'add' from add.cg */
    CGprogram program = cgCreateProgramFromFile( context, CG_SOURCE,
                        "add.cg", profile, "add", 0 );

    /* load the program */
    cgGLLoadProgram( program );

    /* create frame buffer object */
    GLuint fb;
    glGenFramebuffersEXT( 1, &fb );
    glBindFramebufferEXT( GL_FRAMEBUFFER_EXT, fb );
    
    /* create texture ids for x, y, z */
    GLuint tex_x, tex_y, tex_z;
    glGenTextures( 1, &tex_x );
    glGenTextures( 1, &tex_y );
    glGenTextures( 1, &tex_z );

    /* bind texture x so all further texture related calls refer to it */
    glBindTexture( GL_TEXTURE_RECTANGLE_ARB, tex_x );
    
    /* set texture parameters for x */
    glTexParameteri( GL_TEXTURE_RECTANGLE_ARB, GL_TEXTURE_MIN_FILTER, GL_NEAREST);
    glTexParameteri( GL_TEXTURE_RECTANGLE_ARB, GL_TEXTURE_MAG_FILTER, GL_NEAREST);
    glTexParameteri( GL_TEXTURE_RECTANGLE_ARB, GL_TEXTURE_WRAP_S, GL_CLAMP);
    glTexParameteri( GL_TEXTURE_RECTANGLE_ARB, GL_TEXTURE_WRAP_T, GL_CLAMP);
    
    /* actually create texture x */
    glTexImage2D( GL_TEXTURE_RECTANGLE_ARB, 0, GL_FLOAT_RGBA32_NV,
                  M, N, 0, GL_RGBA, GL_FLOAT, 0 );
    
    /* attach texture x so that we can transfer data to it */
    glFramebufferTexture2DEXT( GL_FRAMEBUFFER_EXT, GL_COLOR_ATTACHMENT0_EXT,
                               GL_TEXTURE_RECTANGLE_ARB, tex_x, 0 );
    
    /* transfer array x to texture x */
    glTexSubImage2D( GL_TEXTURE_RECTANGLE_ARB, 0, 0, 0, M, N, GL_RGBA, GL_FLOAT, x );

    /* do everything for y as well */
    glBindTexture( GL_TEXTURE_RECTANGLE_ARB, tex_y );
    glTexParameteri( GL_TEXTURE_RECTANGLE_ARB, GL_TEXTURE_MIN_FILTER, GL_NEAREST);
    glTexParameteri( GL_TEXTURE_RECTANGLE_ARB, GL_TEXTURE_MAG_FILTER, GL_NEAREST);
    glTexParameteri( GL_TEXTURE_RECTANGLE_ARB, GL_TEXTURE_WRAP_S, GL_CLAMP);
    glTexParameteri( GL_TEXTURE_RECTANGLE_ARB, GL_TEXTURE_WRAP_T, GL_CLAMP);
    glTexImage2D( GL_TEXTURE_RECTANGLE_ARB, 0, GL_FLOAT_RGBA32_NV,
                  M, N, 0, GL_RGBA, GL_FLOAT, 0 );
    glFramebufferTexture2DEXT( GL_FRAMEBUFFER_EXT, GL_COLOR_ATTACHMENT0_EXT,
                               GL_TEXTURE_RECTANGLE_ARB, tex_y, 0 );
    glTexSubImage2D( GL_TEXTURE_RECTANGLE_ARB, 0, 0, 0, M, N, GL_RGBA, GL_FLOAT, y );

    /* do everything for z as well, except for transfering data to it */
    glBindTexture( GL_TEXTURE_RECTANGLE_ARB, tex_z );
    glTexParameteri(GL_TEXTURE_RECTANGLE_ARB,GL_TEXTURE_MIN_FILTER,GL_NEAREST);
    glTexParameteri(GL_TEXTURE_RECTANGLE_ARB,GL_TEXTURE_MAG_FILTER,GL_NEAREST);
    glTexParameteri(GL_TEXTURE_RECTANGLE_ARB,GL_TEXTURE_WRAP_S,GL_CLAMP);
    glTexParameteri(GL_TEXTURE_RECTANGLE_ARB,GL_TEXTURE_WRAP_T,GL_CLAMP);
    glTexImage2D( GL_TEXTURE_RECTANGLE_ARB, 0, GL_FLOAT_RGBA32_NV,
                  M, N, 0, GL_RGBA, GL_FLOAT, 0 );
  
    /* specify the first argument x of the program 'add' */
    cgGLSetTextureParameter( cgGetNamedParameter( program, "x" ), tex_x );
    cgGLEnableTextureParameter( cgGetNamedParameter( program, "x" ) );
    
    /* specify the second argument y of the program 'add' */
    cgGLSetTextureParameter( cgGetNamedParameter( program, "y" ), tex_y );
    cgGLEnableTextureParameter( cgGetNamedParameter( program, "y" ) );

    /* specify the outgoing texture to be z */
    glFramebufferTexture2DEXT( GL_FRAMEBUFFER_EXT, GL_COLOR_ATTACHMENT0_EXT,
                               GL_TEXTURE_RECTANGLE_ARB, tex_z, 0 );
    glDrawBuffer( GL_COLOR_ATTACHMENT0_EXT );
   
    /* tell Cg to use our profile and program */
    cgGLEnableProfile( profile );
    cgGLBindProgram( program );
   
    /* run the program */
    glBegin( GL_QUADS );
    {
        glVertex2f( -M, -N );
        glVertex2f( M, -N );
        glVertex2f( M, N );
        glVertex2f( -M, N );
    }
    glEnd(  );
    
    /* read back result from texture z to array z */
    glFramebufferTexture2DEXT( GL_FRAMEBUFFER_EXT, GL_COLOR_ATTACHMENT0_EXT,
                               GL_TEXTURE_RECTANGLE_ARB, tex_z, 0 );
    glReadBuffer( GL_COLOR_ATTACHMENT0_EXT );
    glReadPixels( 0, 0, M, N, GL_RGBA, GL_FLOAT, z );

    /* now the result of z = x + y is pointed to by pointer z */
    /* let's see the first couple of terms and compare with the result on the CPU */
    printf( "\nThese should be really equal, the first column is\n" );
    printf( "the result from the GPU the second is from the CPU:\n\n" );
    for( i = 0; i < 10; i++ )
        printf( "%f = %f\n", z[i], x[i] + y[i] );
    
    /* do the clean up */
    free( x );
    free( y );
    free( z );
    glDeleteTextures( 1, &tex_x );
    glDeleteTextures( 1, &tex_y );
    glDeleteTextures( 1, &tex_z );
    glDeleteFramebuffersEXT( 1, &fb );

    return 0;
}
\end{verbatim}
}

Once both {\tt add.cg} and {\tt add.c} are ready {\tt add.c} can be compiled in the usual way and then should be linked with the
following options {\tt -lCg -lCgGL -lGL -lglut -lGLEW -lpthread}. The resulting executable should be run from the directory
where {\tt add.cg} is located. If no error occurs the output will indicate that indeed the calculation on the GPU and on the
CPU is the same.

\section*{Acknowledgments}

We would like to thank Dominik G\"oddeke for enlightening discussions on the ins and outs
of general purpose GPU programming. In addition L\'aszl\'o Szirmay-Kalos and Tam\'as Umenhoffer are kindly acknowledged
for discussions in the early stages that greatly helped kickstarting our project.
We are grateful to both Domniki G\"oddeke and Julius Kuti for carefully reading the manuscript.

% \newpage

% \section*{Appendix}

% \begin{table}[!h]
% \centering
% \begin{tabular}{ccc}
% \hline
% volume      &     7900 GTX        & 7800 GTX   \\
% \hline
% \,          &  \,              &   \,       \\
% $8^4$       &       3.7        &   3.8      \\
% $8^3 16$    &       7.13       &   7.3      \\
% $8^3 32$    &       10.7       &   8.77     \\
% $16^3 8$    &       10.8       &   8.92     \\
% $16^4$      &       10.9       &   9.03     \\
% $16^3 24$   &       11.3       &   9.19     \\
% $16^3 36$   &       11.5       &   \,        \\
% $24^3 16$   &       11.6       &   \,        \\
% $16^3 60$   &       11.6       &   \,       \\
% \hline
% \end{tabular}
% \caption{Sustained performance in GFlops of the NVIDIA 7900 GTX and 7800 GTX cards measured by the Wilson kernel.}
% \end{table}

\end{document}